\documentclass[12pt
]{article}%

\usepackage{settings}

\begin{document}

\title{\vspace{-0mm} The debt aversion survey module: An experimentally validated tool to measure individual debt aversion%
\footnote{
Corresponding author: \href{mailto:david.albrecht@pm.me}{david.albrecht@pm.me}. 
Thomas Meissner acknowledges funding from the European Union's Horizon 2020 research and innovation programme under grant agreement No. 795958. Replication files can be found via OSF: \url{https://osf.io/xzhw4/?view_only=385a9542dd11478286ca43e79c2f891c}}
}
\author{
\normalsize\begin{tabular}{c}
David Albrecht\footnote{Maastricht University} \\
Thomas Meissner\footnotemark[2]\\
[2ex]
\end{tabular}
}
\maketitle
\bigskip
\bigskip

\renewcommand{\baselinestretch}{1} \normalsize
\begin{abstract}
\noindent We develop an experimentally validated, short and easy-to-use survey module for measuring individual debt aversion. To this end, we first estimate debt aversion on an individual level, using choice data from \cite{meissner2022}. This data also contains responses to a large set of debt aversion survey items, consisting of existing items from the literature and novel items developed for this study. Out of these, we identify a survey module comprising two qualitative survey items to best predict debt aversion in the incentivized experiment. 

\keywords{Debt Aversion \and Preference Measurement \and Survey Validation}
\end{abstract}
\renewcommand{\baselinestretch}{1.45} \normalsize
\setlength{\footnotesep}{0.75\baselineskip}

\newpage
\section{Introduction}

In this paper, we develop an experimentally validated, short and easy-to-use survey module for measuring individual debt aversion. Debt aversion has been shown to affect financial decision-making in a variety of different contexts, such as financing (tertiary) education \citep{field2009, caetano2019}, private investments in real estate \citep{schleich2021} or entrepreneurial activity \citep{nguyen2020}. In \cite{meissner2022} we show that debt aversion has a strong impact on borrowing and saving decisions and that it captures a genuine preference that is distinct from other individual preferences for time, risk and losses. However, while debt aversion appears to have a strong influence on financial decision-making, it is notoriously difficult to measure, typically involving complex and resource-intensive experiments \citep[see e.g.][]{Meissner2016,ahrens2022intertemporal,meissner2022}. This complicates the study of debt aversion on a larger scale and impedes the understanding of debt aversion and its effects in representative populations. 

In this paper, we develop an experimentally validated, short and easy-to-use survey module for measuring individual debt aversion. Using the choice data and theoretical framework of \cite{meissner2022}, we structurally estimate parameters of debt aversion on the individual level, using hierarchical maximum likelihood methods. Building on the methodology of the Global Preference Survey - GPS \citep{falk2018globalpreferences, falk2022surveymodule} we then test the predictive power of a large set of debt aversion survey items, consisting of existing items from the literature and novel items developed for this study. Out of these, we select a smaller subset of survey items, that jointly best predict debt aversion in the incentivized experiment as our debt aversion survey module.

The module that provides the best trade-off between in- and out-of-sample fit, as well as brevity of implementation, consists of two Likert scale survey items, summarized in Table \ref{table:fullmodule}. This survey module is short, easy to implement and predicts debt aversion reasonably well. It may therefore facilitate future work on debt aversion, in research where expensive and complex incentivized experiments are not feasible, such as large-scale representative surveys. Further, the survey module might also improve behavioral predictions and identification in research where debt aversion is not the main focus.  In such contexts, it enables to easily control for individual preferences towards debt and thus may prevent inference on spurious relationships.

In the following, we first introduce the debt aversion survey module in Section \ref{sec:survey_module} and delineate its practical implementation in detail. Next, we summarize the design and procedures of the validation study in Section \ref{sec:validation_study} which provides experimentally measured individual preferences on risk, time and losses as well as debt aversion as the base of the survey module development. Section~\ref{sec:modselection} outlines the details of identifying the debt aversion survey module. Finally, in Section~\ref{sec:discussion} we discuss the strengths and limitations of the identified survey module.

\FloatBarrier
\section{Debt Aversion Survey Module}
\label{sec:survey_module}
Balancing predictive accuracy and brevity of implementation, we propose the debt aversion survey module in Table~\ref{table:fullmodule}. It consists of two qualitative survey items: \emph{Q1} and \emph{Q2}. Table~\ref{table:fullmodule} provides the exact wording of questions and the corresponding response categories, which can be directly implemented in any questionnaire as part of an experiment or a survey.\footnote{Note that item \emph{Q2} asks about second-order beliefs and is thus introduced differently than \emph{Q1}.}

As in \cite{meissner2022}, we denote $\gamma$ as the parameter of debt aversion. $\gamma=1$ implies debt neutrality, $\gamma>1$ implies debt aversion and $\gamma<1$ implies debt affinity.\footnote{For more details on the underlying theory, please see Appendix \ref{sec:model}.} To translate responses to the survey module into a quantitative measure of debt aversion, the coefficients in the last column of Table~\ref{table:fullmodule} can be used as weights to construct $\gamma$, as the weighted sum of the module's items and the constant. These weights represent estimated OLS coefficients stemming from the multivariate regression of the debt aversion parameter $\gamma$, the dependent variable, on the survey module's set of predictors as independent variables. Likert Scale ratings are coded in the range of 1 to 6 corresponding to answers from ``strongly agree" to ``strongly disagree''. 
We include a hypothetical example in Appendix~\ref{sec:calc_example}.

\begin{table}[t!]
\caption{Debt Aversion Survey Module \label{table:fullmodule}}
\begin{tabularx}{\textwidth}{
    c 
    L{7cm}
    C{3.5cm}
    >{\centering\arraybackslash}X 
}

\toprule
& \textbf{Question/Item} 
& \textbf{Response} 
& \textbf{Coeff.} \\
\midrule

\multicolumn{2}{l}{\footnotesize{Please rate the following statement:}}
& 
&\\[+4pt]

\footnotesize{\emph{Q1}}
&\footnotesize{\emph{Debt is an integral part of today's life.}}
& \footnotesize{1:~strongly agree - 6:~strongly disagree}
& \footnotesize{$0.0045$} \\[+16pt]


\cline{1-2}
\multicolumn{2}{l}{\footnotesize{What do you think, how does the average participant in  }}
&
&\\[-4pt]
\multicolumn{2}{l}{\footnotesize{this survey/experiment rate the following statement: }}
& 
&\\[+4pt]

\footnotesize{\emph{Q2}}
&\footnotesize{\emph{There is no excuse for borrowing money.}}
& \footnotesize{1:~strongly agree - 6:~strongly disagree}
& \footnotesize{$-0.0067$} \\[+16pt]



\cline{1-2}

\footnotesize{\textit{Constant}} 
& 
& 
& \footnotesize{$1.0694$} \\

\bottomrule


\end{tabularx}
\end{table}

The survey module's quality can be quantified by several metrics. 
With respect to within-sample fit, the survey module and the parameter of debt aversion elicited through the incentivized experiment reach a correlation of $0.3073$.\footnote{Preferences over risk, time and losses are not significantly correlated with debt aversion as measured by the survey module.} This value is close to the lower end of in-sample correlations of the GPS survey module \citep{falk2022surveymodule}, which range from 0.37 to 0.67. Beyond in-sample fit, k-fold cross-validation yields an average mean absolute prediction error (MAE) of $0.0272$.\footnote{The MAE is calculated based on 100 repeated random samples with k=5 and k=10 partitions of the data, respectively. The reported figure is the average mean absolute error over the two levels of $k$.} In other words, predicting the individual debt aversion parameter with the survey module is subject to an average imprecision of $0.0272$ points. 
Evaluated in isolation, these quality metrics might understate the actual power of the survey module. In Section~\ref{sec:discussion}, we set the metrics into perspective by comparing them to suitable benchmarks in particular and by discussing the virtues and limitations of the survey module in general.

\section{Validation Study}
\label{sec:validation_study}

To construct the debt aversion survey module we make use of the theoretical framework and experimental data from \cite{meissner2022}. Both taken together enable the quantification of debt aversion on the individual level. A summary of the theory of debt aversion can be found in Appendix \ref{sec:model}. For a more detailed description of the theory and corresponding experiment, we refer the reader to the original paper. We use the experimental choice data to structurally estimate the parameter of debt aversion~$\gamma$ on the individual level. 
Using the individual estimates of $\gamma$, we can construct a debt aversion survey module that optimally predicts~ $\gamma$ and thus the degree of debt aversion. The following paragraphs provide a summary of the experiment and detail the procedures to structurally estimate preference parameters on the individual level. 

\subsection{Preference Elicitation and Estimation}
\label{sec:elicitation_and_estimation}

A total of 127 people completed a three-session sequence of experiments in the behavioral and experimental laboratory BEElab at Maastricht University conducted over the years 2019 to 2021. Important for this work, they decided on a total of 90 
binary prospects defined over payments at different points in time, lotteries, as well as saving and debt contracts. These choices can be grouped in seven 
different multiple price lists (MPLs) which are presented in Appendix~\ref{sec:mpls}.

 To construct the debt aversion survey module, we extend the estimation procedure of \cite{meissner2022} to additionally estimate preference parameters on the individual level. In \cite{meissner2022} we use the entirety of all 90 choices made by all participants to jointly estimate preference parameters of risk, loss and debt aversion as well as time discounting, using maximum likelihood procedures. We estimate preferences on the aggregate level, i.e. for the average decision-maker. Additionally, we estimate moments of the population distributions of preference parameters. Extending these procedures for this study, we employ hierarchical maximum likelihood estimation to retrieve individual preference parameters, loosely following \cite{murphy2018} and \cite{farrell2008}. This technique has been established to increase the reliability of individual preference parameter estimates which have to rely on far less data than aggregate estimations. Hierarchical maximum likelihood procedures estimate the set of individual preference parameters that is most likely to produce the observed choices of the respective person, weighted by the probability of occurrence of such parameter estimates given the population distribution of parameter estimates.

Based on the random utility model outlined in \cite{meissner2022}, a decision maker with preference parameters $\omega=(\alpha,\beta,\gamma,\lambda)$ chooses option B if $U(X^B,\omega)+\varepsilon^{B}\geq U(X^A,\omega) +\varepsilon^{A}$. The probability of observing choice B can then be written as:

\begin{align}
\label{eq:cond_prob}
P^B(\theta)& = F\left(\frac{U(X^B,\omega)-U(X^A,\omega)}{\mu}\right) = F(\Delta U(\theta)),
\end{align}

where $F$ is the cumulative distribution function of $(\varepsilon^A-\varepsilon^B)$. We assume $(\varepsilon^A-\varepsilon^B)$ to follow a standard logistic distribution with distribution function $F(\xi)=(1+\mathrm{e}^{-\xi})^{-1}$, corresponding to the often termed Luce model \citep{luce1965preference, holt2002} or Fechner error with logit link \citep{drichoutis2014}. $F(\Delta U(\theta))$ depends on the preference parameters for risk aversion, time discounting, debt aversion and loss aversion: $\alpha$, $\delta$, $\gamma$ and $\lambda$, respectively and the error parameter $\mu$. All five parameters, summarized by the vector $\theta=(\alpha,\beta,\gamma,\lambda,\mu)$, are identified on the individual level through hierarchical maximum likelihood estimation based on observed choices and the population distribution of preference parameters.\footnote{In line with \cite{murphy2018} we do not consider the population distribution of the error parameter $\mu$ for hierarchical estimation, as it cannot be estimated and interpreted independently of other preference parameters.} The population distributions have already been retrieved in the first hierarchy level of estimation in \cite{meissner2022}. Thus, we here consider them as given by the normal density functions $d(\alpha)$, $d(\delta)$, $d(\gamma)$ and $d(\lambda)$. We denote the product of these normal densities as $d(\omega)=d(\alpha)d(\delta)d(\gamma)d(\lambda)$. The hierarchical likelihood function on the individual level then writes as

\begin{align}
\label{eq:hlogL1}
\begin{split}
ln \left( L(\theta) \right) = 
\sum_j\left[ln\left(F(\Delta U(\theta))w(d(\omega))\right)c_{j}+ln\left(1-F(\Delta U(\theta))w(d(\omega))\right)(1-c_{j})\right],
\end{split}
\end{align}

where $c_{j}=0$ if Option~A was chosen in choice~$j$, $c_{j}=1$ if Option~B was chosen in choice~$j$. Further, $w(d(\omega))$ denotes the product of preference parameters' density functions, given their distribution in the population, weighted by function $w(\cdot)$. 

We introduce $w$ to prevent a potential pitfall of hierarchical estimation. \cite{scheibehenne2015} argue that the relative influence of population distributions on individual preference parameters, as opposed to the influence of individual choices, may be too large. In other words, hierarchical estimation may produce excessive shrinkage of individual parameter estimates toward their population mean.\footnote{We find such excessive shrinkage to likely also apply to our setting, based on estimations for simulated decision makers, with hypothetical, known preference parameters.}
Following a suggestion by \cite{murphy2018} we mitigate this effect by weighing $d(\omega)$. In particular, we introduce the weighting function

\begin{align}
\label{eq:weighting_function}
    w(d(\omega))=d(\omega)^s,
\end{align}

where $s$ is a shrinkage factor determining the relative weight given to population distributions as opposed to individual choices when estimating individual preference parameters.\footnote{We employ the exponential form, as it preserves the range of $d(\omega)$, i.e. both $d(\omega)$ and $w(d(\omega))$ lie in the range of probabilities $[0,1]$. Further, the exponential form introduces only one additional parameter to be estimated.} If $s=1$, then $w(d(\omega))=d(\omega)$, for $s<1$ the influence of population distributions on estimates decreases, and for $s>1$ the influence increases. 

We estimate the optimal shrinkage factor $s$ based on simulated, hypothetical individuals with known preference parameters. To this end, we apply the outlined hierarchical ML procedure to estimate preference parameters from simulated choices. By comparing the estimated parameters to true, assumed parameters we can quantify the goodness-of-fit. Using grid-search, we find $s=0.0139$ 
to minimize mean squared error (MSE) of estimated and true individual parameters of debt aversion $\gamma$.


\begin{figure}
\caption{Individual level estimates of the debt aversion parameter $\gamma$. \label{figure.indiv_parameter_freq}}
\centering
\begin{tikzpicture}
\begin{axis}[
	width  = 0.8*\textwidth,
	height = 5cm,
	ybar,
	ymin=0,
	ymax=28,
	axis lines = left,
	ylabel = no. of individuals,
	ymajorgrids = true,
	xmin = 0.9,
	xmax = 1.175,
	xlabel= debt aversion ($\gamma$),
	xtick = data,
	extra x ticks={1},
	extra x tick style={xticklabel=1},
	xticklabels={,,0.95,,1,,1.05,,1.1,,1.15,},
	area style,
]

\addplot+[ybar interval,mark=no,] plot coordinates {	
                                                    (0.9, 0) 
													(0.925, 1)
													(0.95, 1)
													(0.975, 2)
													(1.0, 16)
													(1.025, 24)
													(1.05, 24)
												    (1.075, 20)
													(1.1, 8)
													(1.125, 0)
													(1.15, 0)
													};

\end{axis}
\end{tikzpicture}

\end{figure}
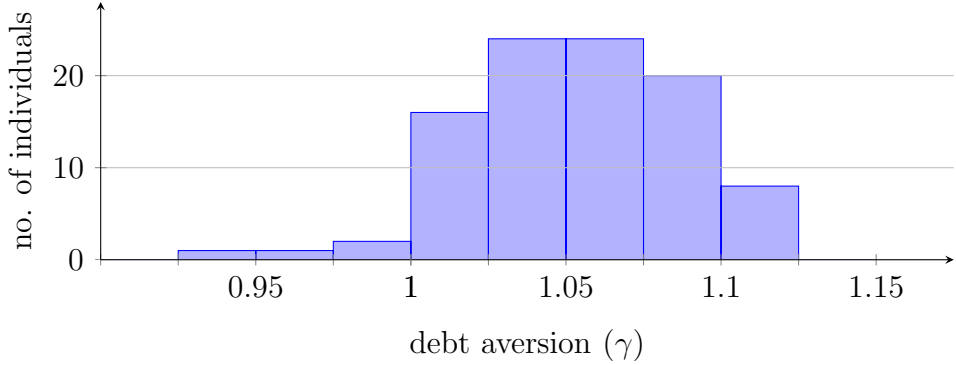

By maximizing the hierarchical log-likelihood function (Equation~\ref{eq:hlogL1}) based on choice data and population distributions from \cite{meissner2022} we derive point estimates for all preference parameters and the error parameter per individual.\footnote{To increase the reliability of estimates we ensure that the same estimates can be retrieved using STATA's modified Broyden-Fletcher-Goldfarb-Shanno (BFGS) algorithm and through maximization using the Nelder-Mead algorithm implemented in R. Individuals, whose parameters cannot be estimated consistently, i.e. estimates from R are more than $10\%$ higher or lower than estimates from STATA, are discarded. For consistent estimates, we choose results from either STATA or R depending on which routine yielded a higher likelihood score for the given individual's parameter estimates.} In this way, we retrieve individual preference parameters, including $\gamma$, the coefficient of debt aversion, for 96 participants. Figure~\ref{figure.indiv_parameter_freq} summarizes the distribution of $\gamma$. Individual parameter estimates range from 
$\gamma=0.9468$, corresponding to debt affinity, to $\gamma=1.1171$, corresponding to debt aversion. Around, 96\% of individuals are estimated as debt averse with a parameter $\gamma>1$, the median level of debt aversion is $\gamma=1.0523$.

\subsection{Included survey items}
\label{sec:EoIC_ed}
Besides incentivized choices, we collected a battery of self-reported, qualitative measures of debt behavior and attitudes in \cite{meissner2022}. For this purpose, we composed a comprehensive set of 54 survey items, roughly spanning the clusters experience and usage of borrowing, appropriateness to be indebted, rules, norms and personal preferences on debt. The items are collected from previous studies using questionnaire-based measures of debt attitudes but also include novel items developed for this study. Items range from directly asking respondents to state ``how much they (dis)like being in debt in general'' to more indirect survey items around the topic of debt and money. The complete list of items can be found in Appendix, Table~\ref{table:debtsurvey_all}.
In addition to the set of survey items, we implemented a hypothetical, multiple price list consisting of 15 debt contracts. To minimize the required response time, the hypothetical debt contracts were presented using the staircase method \citep{RN593}. This mode of presentation allows to identify the switchpoint between accepting and not accepting each of the 15 contracts by only asking four successive questions. for detailed instructions see Appendix~\ref{sec:DSM_HDC_a}. We count the switchpoint in the hypothetical choice task as one additional quantitative survey item.

\section{Identification of the Debt Aversion Survey Module}
\label{sec:modselection}
To identify an experimentally validated survey module for debt aversion we follow the procedures established for the GPS \citep{falk2022surveymodule}. To construct the survey module, we first consider the entirety of collected survey items, and all possible combinations thereof. We then identify the subset of items that yields the best combination of predictive accuracy for individual debt aversion and brevity. 
We begin by considering a baseline pool of 55 potential best predictors including qualitative items collected from existing literature on debt attitudes, novel items developed for this study and the switchpoint in the hypothetical debt contract choice task. By stepwise reduction, we condense the amount of predictors to a debt aversion survey module of two items.              
                
In the first step, we discard thirteen items that appear improper to construct an easy and widely applicable survey module. Six items have a specific focus on (debt) financing tertiary education, which makes them inappropriate for use beyond the university context. For further five items, we could not identify an intuitive directional hypothesis on the relation of the item and debt aversion. Lastly, two items exhibit a correlation with debt aversion, which goes against the direction of an intuitive hypothesis. A potential reason for this could be a misunderstanding of these items due to double negative wording.

As a second step, we consider a multitude of linear regressions, modelling the experimentally elicited debt aversion parameter $\gamma$ per individual $i$ as the dependent variable, and all possible subsets of the remaining $42$ items as independent variables. In other words, we scrutinize all possible combinations of one item, two items, ..., $n$ items as independent variables in a standard linear model, see Equation~\ref{eq:regression}, and estimate the regression parameters $\beta_0,...,\beta_p$ using ordinary least squares (OLS).

\begin{align}
    \gamma_i=\beta_0+\beta_1x_{1i}+...+\beta_px_{pi}+\varepsilon_i
    \label{eq:regression}
\end{align}

The number of potential combinations, increases rapidly in the number of considered items, and quickly becomes intractable with conventional computational resources.\footnote{
$42$ combinations of a single predictor, i.e. each and every item itself. $861$ combinations of two predictors, $11\,480$ combinations of three predictors, $111\,930$ combinations of four predictors, and so on, with a maximum of $538\,257\,874\,440$ combinations of $21$ predictors. As subsets/combinations of predictors are sampled disregarding the order of elements and without the possibility to include the same element more than once the number of combinations reaches a maximum for models including $21$ variables. For larger subsets, the number of potential combinations decreases again. There is only one subset containing all predictors from the pool.
}

We therefore create a shortlist of  predictor variables that appear in any of the ten best performing models, evaluated according to adjusted $R^2$ for models with one, two, three up to six predictors, respectively.\footnote{For the creation of this shortlist we run a total of $36\,211\,980$ regressions.}
 This shortlist contains 16 items. Using the shortlist we run linear OLS regressions on all possible combinations of predictors, i.e. $524\,288$ regressions. In line with the GPS module \citep{falk2022surveymodule}, we use adjusted $R^2$, a criterion of in-sample-fit, to identify the best subset for each number of predictors.

\begin{figure}[t!]
\caption{Performance metrics of the candidate survey modules
\label{fig.performance}}

\begin{subfigure}{1\textwidth}
\begin{minipage}{0.49\textwidth}
\begin{tikzpicture}
\centering
    \hspace{9pt}
	\begin{axis}[
		width  = 0.95*\textwidth,
		height = 5.5cm,
		ymin=0,
		ymax=0.25,
		axis lines = left,
		ylabel = \footnotesize{$R^2$},
		ymajorgrids = true,
		xmin = 0,
		xmax = 13,
		tick label style={font=\footnotesize},
		legend entries={\scriptsize{Top1}, \scriptsize{Top2}, \scriptsize{Top3},
        \scriptsize{Top4}, \scriptsize{Top5}},
        legend style={at={(0.975,0.65)}},
		]
        \addplot [red,very thick] table [x="n_tuple", 
                                        y="r.squared.1", 
                                        col sep=comma, 
                                        mark=none] {RESULTS/survey_module_selection.csv};
        \addplot [darkgray] table [x="n_tuple", 
                               y="r.squared.2",
                               col sep=comma, 
                               mark=none] {RESULTS/survey_module_selection.csv};
        \addplot [gray] table [x="n_tuple", 
                               y="r.squared.3",
                               col sep=comma, 
                               mark=none] {RESULTS/survey_module_selection.csv};
        \addplot [gray, densely dotted] table [x="n_tuple",
                               y="r.squared.4", 
                               col sep=comma, 
                               mark=none] {RESULTS/survey_module_selection.csv};
        \addplot [lightgray, dotted] table [x="n_tuple",
                               y="r.squared.5",
                               col sep=comma, 
                               mark=none] {RESULTS/survey_module_selection.csv};
	\end{axis}
\end{tikzpicture}
\end{minipage}
\begin{minipage}{0.49\textwidth}
\begin{tikzpicture}
\centering
    \hspace{9pt}
	\begin{axis}[
		width  = 0.95*\textwidth,
		height = 5.5cm,
		ymin=0,
		ymax=0.25,
		axis lines = left,
		ylabel = \footnotesize{adjusted $R^2$},
		ymajorgrids = true,
		xmin = 0,
		xmax = 13,
		tick label style={font=\footnotesize},
		]
        \addplot [red,very thick] table [x="n_tuple", 
                                        y="adj.r.squared.1", 
                                        col sep=comma, 
                                        mark=none] {RESULTS/survey_module_selection.csv};
        \addplot [darkgray] table [x="n_tuple",
                                        y="adj.r.squared.2", 
                                        col sep=comma, 
                               mark=none] {RESULTS/survey_module_selection.csv};
        \addplot [gray] table [x="n_tuple", 
                                        y="adj.r.squared.3", 
                                        col sep=comma, 
                               mark=none] {RESULTS/survey_module_selection.csv};
        \addplot [gray, densely dotted] table [x="n_tuple",
                                        y="adj.r.squared.4", 
                                        col sep=comma, 
                               mark=none] {RESULTS/survey_module_selection.csv};
        \addplot [lightgray, dotted] table [x="n_tuple",
                                        y="adj.r.squared.5", 
                                        col sep=comma, 
                               mark=none] {RESULTS/survey_module_selection.csv};
       \addplot[blue, ultra thick, draw opacity=0.1]coordinates{(9,0) (9,0.142844833311355)};
	\end{axis}
\end{tikzpicture}
\end{minipage}
\end{subfigure}

\vspace{1cm}
\begin{subfigure}{1\textwidth}
\begin{minipage}{0.49\textwidth}
\begin{tikzpicture}
\centering
	\begin{axis}[
		width  = 0.95*\textwidth,
		height = 5.5cm,
		ymin=-658,
		ymax=-647,
		axis lines = left,
		ylabel = \footnotesize{AIC},
		ymajorgrids = true,
		xmin = 0,
		xmax = 13,
		tick label style={font=\footnotesize}
		]
        \addplot [red,very thick] table [x="n_tuple",
                                        y="AIC.1", 
                                        col sep=comma, 
                                        mark=none] {RESULTS/survey_module_selection.csv};
        \addplot [darkgray] table [x="n_tuple",
                                        y="AIC.2", 
                                        col sep=comma, 
                                        mark=none] {RESULTS/survey_module_selection.csv};
        \addplot [gray] table [x="n_tuple",
                                        y="AIC.3", 
                                        col sep=comma, 
                                        mark=none] {RESULTS/survey_module_selection.csv};
        \addplot [gray, densely dotted] table [x="n_tuple", 
                                        y="AIC.4", 
                                        col sep=comma, 
                                        mark=none] {RESULTS/survey_module_selection.csv};
        \addplot [lightgray, dotted] table [x="n_tuple", 
                                        y="AIC.5", 
                                        col sep=comma, 
                                        mark=none] {RESULTS/survey_module_selection.csv};
        \addplot[blue, ultra thick, draw opacity=0.1]coordinates{(6,-661) (6,-656.477233123315)};
	\end{axis}
\end{tikzpicture}
\end{minipage}
\begin{minipage}{0.49\textwidth}
\begin{tikzpicture}
\centering
	\begin{axis}[
		width  = 0.95*\textwidth,
		height = 5.5cm,
		ymin=-652,
		ymax=-608,
		axis lines = left,
		ylabel = \footnotesize{BIC},
		ymajorgrids = true,
		xmin = 0,
		xmax = 13,
		tick label style={font=\footnotesize}
		]
        \addplot [red,very thick] table [x="n_tuple", 
                                        y="BIC.1", 
                                        col sep=comma, 
                                        mark=none] {RESULTS/survey_module_selection.csv};
        \addplot [darkgray] table [x="n_tuple", 
                                        y="BIC.2", 
                                        col sep=comma, 
                                        mark=none] {RESULTS/survey_module_selection.csv};
        \addplot [gray] table [x="n_tuple", 
                                        y="BIC.3", 
                                        col sep=comma, 
                                        mark=none] {RESULTS/survey_module_selection.csv};
        \addplot [gray, densely dotted] table [x="n_tuple", 
                                        y="BIC.4", 
                                        col sep=comma, 
                                        mark=none] {RESULTS/survey_module_selection.csv};
        \addplot [lightgray, dotted] table [x="n_tuple", 
                                        y="BIC.5", 
                                        col sep=comma, 
                                        mark=none] {RESULTS/survey_module_selection.csv};
        \addplot[blue, ultra thick, draw opacity=0.1]coordinates{(1,-652) (1,-647.056288196309)};
	\end{axis}
\end{tikzpicture}
\end{minipage}
\end{subfigure}
\vspace{1cm}

\begin{subfigure}{1\textwidth}
\begin{minipage}{0.49\textwidth}
\begin{tikzpicture}
\centering
	\begin{axis}[
		width  = 0.95*\textwidth,
		height = 5.5cm,
		ymin=0.00105,
		ymax=0.00135,
		axis lines = left,
		ylabel = \footnotesize{MSE (CV with k=5)},
		ymajorgrids = true,
		xmin = 0,
		xmax = 13,
		xlabel= No. of predictors,
		scaled ticks=false, 
		tick label style={font=\footnotesize},
		y tick label style={/pgf/number format/.cd,
        fixed,
        fixed zerofill,
        precision=4,
        yticklabels={,,.0011,.0012,.0013,,},
        /tikz/.cd
        }
		]
        \addplot [red,very thick] table [x="n_tuple", 
                                        y="MSPE_k5_rd100.1", 
                                        col sep=comma, 
                                        mark=none] {RESULTS/survey_module_selection.csv};
        \addplot [darkgray] table [x="n_tuple", 
                                        y="MSPE_k5_rd100.2", 
                                        col sep=comma, 
                                        mark=none] {RESULTS/survey_module_selection.csv};
        \addplot [gray] table [x="n_tuple", 
                                        y="MSPE_k5_rd100.3", 
                                        col sep=comma, 
                                        mark=none] {RESULTS/survey_module_selection.csv};
        \addplot [gray, densely dotted] table [x="n_tuple", 
                                        y="MSPE_k5_rd100.4", 
                                        col sep=comma, 
                                        mark=none] {RESULTS/survey_module_selection.csv};
        \addplot [lightgray, dotted] table [x="n_tuple", 
                                        y="MSPE_k5_rd100.5", 
                                        col sep=comma, 
                                        mark=none] {RESULTS/survey_module_selection.csv};
        \addplot[blue, ultra thick, draw opacity=0.1]coordinates{(2,0.00110130341633874) (2,0.00105)};
	\end{axis}
\end{tikzpicture}
\end{minipage}
\begin{minipage}{0.49\textwidth}
\begin{tikzpicture}
\centering
	\begin{axis}[
		width  = 0.95*\textwidth,
		height = 5.5cm,
		ymin=0.00105,
		ymax=0.00135,
		axis lines = left,
		ylabel = \footnotesize{MSE (CV with k=10)},
		ymajorgrids = true,
		xmin = 0,
		xmax = 13,
		xlabel= No. of predictors,
		scaled ticks=false, 
		tick label style={font=\footnotesize},
		y tick label style={/pgf/number format/.cd,
        fixed,
        fixed zerofill,
        precision=4,
        yticklabels={,,.0011,.0012,.0013,,},
        /tikz/.cd
        }
		]
        \addplot [red,very thick] table [x="n_tuple", 
                                        y="MSPE_k10_rd100.1", 
                                        col sep=comma, 
                                        mark=none] {RESULTS/survey_module_selection.csv};
        \addplot [darkgray] table [x="n_tuple", 
                                        y="MSPE_k10_rd100.2", 
                                        col sep=comma, 
                                        mark=none] {RESULTS/survey_module_selection.csv};
        \addplot [gray] table [x="n_tuple", 
                                        y="MSPE_k10_rd100.3", 
                                        col sep=comma, 
                                        mark=none] {RESULTS/survey_module_selection.csv};
        \addplot [gray, densely dotted] table [x="n_tuple", 
                                        y="MSPE_k10_rd100.4", 
                                        col sep=comma,
                                        mark=none] {RESULTS/survey_module_selection.csv};
        \addplot [lightgray, dotted] table [x="n_tuple", 
                                        y="MSPE_k10_rd100.5", 
                                        col sep=comma, 
                                        mark=none] {RESULTS/survey_module_selection.csv};
        \addplot[blue, ultra thick, draw opacity=0.1]coordinates{(2,0.00109987097281651) (2,0.00105)};
	\end{axis}
\end{tikzpicture}
\end{minipage}
\end{subfigure}

\vspace{0.5cm}
\footnotesize

\footnotesize{Notes: \textbf{First row:} coefficients of determination;
\textbf{Second row:} information criteria;
\textbf{Third row:} mean squared error (MSE) derived through k-fold cross validation (CV) with $k=5$ and $k=10$ data chunks and 100 repetitions each}
\end{figure}
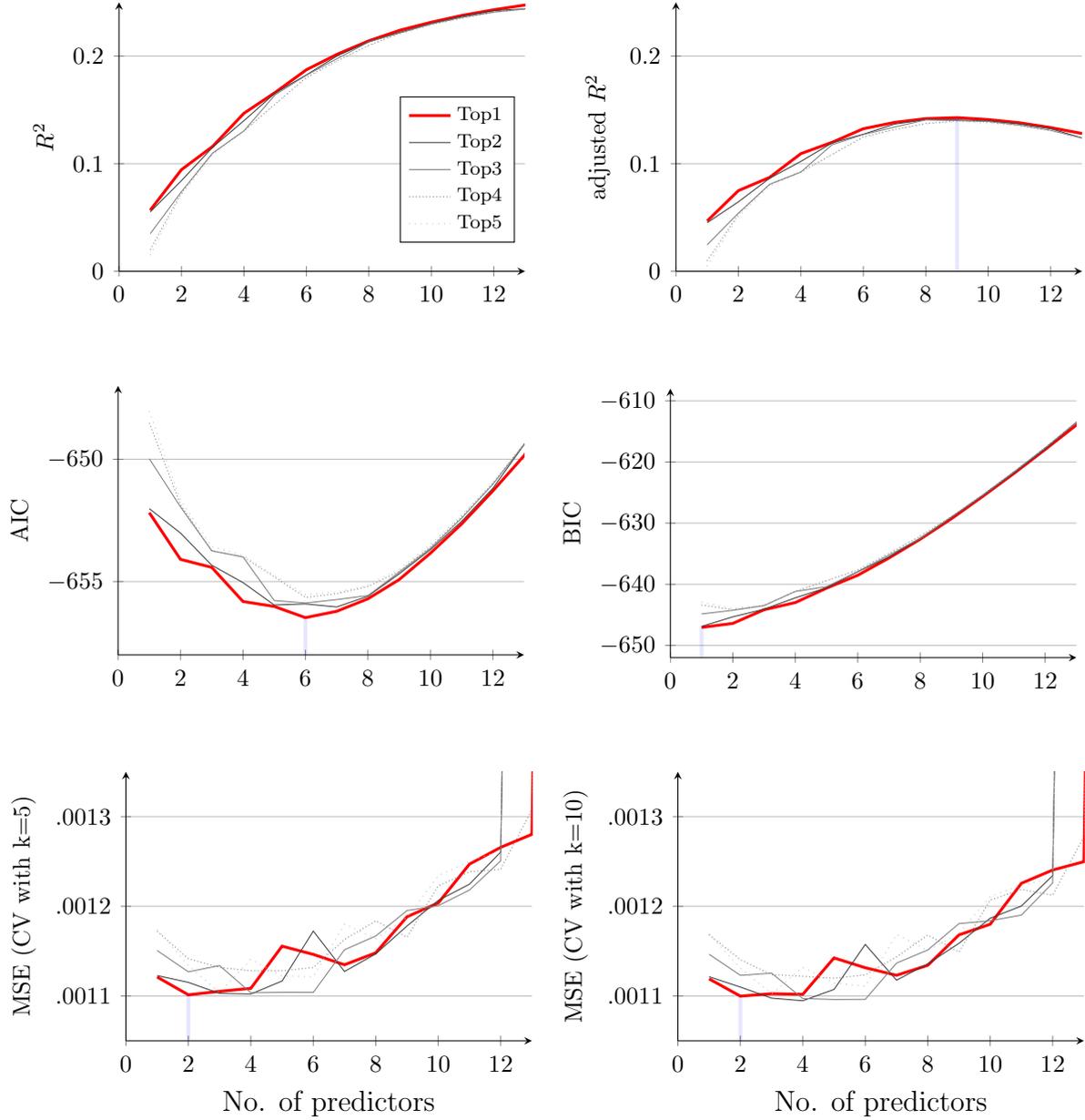

Third, to discriminate between models comprising different numbers of variables we additionally consider information criteria and estimates of out-of-sample predictive power based on cross-validation. We consider the Akaike Information Criterion (AIC) as introduced by \cite{RN619} and the Bayesian or Bayes-Schwarz Information Criterion (BIC) as introduced by \cite{RN620}.
With respect to cross-validation, we implement k-fold cross-validations \citep{RN622} splitting our sample into $k=5$ and $k=10$ data chunks with 100 random samples each to calculate mean squared prediction errors (MSE) for the parameter of debt aversion $\gamma$ of the candidate models. The performance of the five best candidate models per number of predictors are summarized in Figure~\ref{fig.performance}.

Bringing together the variety of performance measures calculated in the previous step, we identify a survey module containing two items as providing a good trade-off between brevity and predictive power.
Figure~\ref{fig.performance} shows that adjusted $R^2$ is maximized for a model with nine predictors, while AIC favors a six and BIC a one predictor model. The discrepancy in terms of favored models does not come as a surprise, as the three performance criteria differ in the penalty they put on including additional predictors into the model. Adjusted $R^2$ incorporates the smallest and BIC the largest penalty. To complement in-sample fit based performance metrics, we scrutinize the cross-validation-based MSE of the candidate models. Successively including more predictors, MSEs for $k=5$ and $k=10$ decrease for the Top1 models with up to two predictors, remain relatively flat up to four predictors and increases for more than four predictors. These favored numbers of predictors are largely corroborated by considering not only the best model (Top1) but also its close competitors (Top2 to Top5).

Just as in the GPS module \citep{falk2022surveymodule} we value brevity of the survey module and thus favor the BIC. However, we do not want to fully disregard adjusted $R^2$ and AIC, as the discrepancy of suggested predictors is rather large. We reconcile the number of predictors by following the best model according to cross validation, comprising two predictors. This appears to be a reasonable compromise between BIC and AIC, which still puts a strong weight on brevity.

\section{Discussion and conclusion}
\label{sec:discussion}
We propose a two-question survey module, that allows measuring debt aversion in large-scale surveys and experimental questionnaires with minimal effort and expenditure of time. The module is validated by use of an incentivized experiment, thus lending credibility to the resulting measure. In the following, we discuss the performance of the debt aversion survey module according to different metrics and in comparison to reasonable benchmarks. This way we provide guidance for researchers who face the trade-off between eliciting debt aversion via the short, but potentially less accurate survey module as opposed to incentivized lab experiments and structural estimations, which are expensive and involve considerable time and effort.\footnote{In \cite{meissner2022}, participants had to come to the lab on a total of three dates, with a total time of around 2.5 hours.}

As a starting point, we consider the in-sample fit of the debt aversion survey module as indicator of quality. The module reaches an $R^2$ of $0.0945$, i.e. it accounts for around 10\% of the variation of the parameter of debt aversion (from the incentivized experiment). This corresponds to a correlation of $\rho= 0.3073$. 
As discussed for the GPS module \citep{falk2022surveymodule}, this correlation cannot be evaluated against a benchmark of 1, unless measurement of preferences (both in the structural estimations and in the response to the survey items) was without measurement error. 

To evaluate measurement error, repeated measurement on the same participants would be helpful. While this was not done by \cite{meissner2022}, the validation study for the GPS module finds a test-retest correlation of $0.5753$ for risk aversion, and $0.8149$ for time discounting when comparing their preference measure in one experiment to the same measure from an identical repeated experiment with the same participants \citep{falk2022surveymodule}. 
Evaluated against such a benchmark, our survey module appears to exhibit reasonable explanatory power for measuring debt aversion.


Beyond in-sample fit, the debt aversion survey module proves to be capable in terms of predictive power. K-fold cross validation yields a mean absolute prediction error (based on 100 repeated random samples) of $0.0272$ for k=5 and  k=10. In other words, predicting $\gamma$ using the debt aversion survey module suffers an average error of ca. $0.0272$\footnote{Calculated as the mean absolute prediction error over the two different $ks$}. 
To set this into perspective we consider predicted choices for participants in the experiment of \cite{meissner2022} in comparison to their actual choices. Specifically, we consider predictions based on the debt aversion parameter from the survey module competing against predictions based on the structurally estimated debt aversion parameter as a benchmark. Survey module based predictions are accurate in $89.76\%$, while structural estimation based predictions are accurate in $91.48\%$ of the decision situations in the experiment. Thus, the survey module performs similarly well in predicting choices compared to the structurally estimated parameter of debt aversion. 

Comparing our results to the module of the GPS \citep{falk2018globalpreferences, falk2022surveymodule}, we see the debt aversion survey module as a useful addition to measure debt aversion alongside other key economic preferences in survey style investigations. A notable difference to the GPS module is that we do not identify the hypothetical version of incentivized choices, in our case the hypothetical debt contracts, as part of the set of the best-predictors of preferences. We see a likely explanation for this in the interdependence of debt aversion with other preference dimensions such as risk and loss aversion as well as discounting over time. As illustrated in \cite{meissner2022}, the switch point in a multiple price list consisting of debt contracts will not only depend on individual debt aversion, but potentially also on time preferences, risk preferences and loss aversion. 

Summing up, we develop a short and easy-to-use experimentally validated  survey module to measure debt aversion. We hope that our survey module will facilitate future research on debt aversion on a larger scale, where complex and incentivized experiments are often not feasible. 

\newpage

\bibliographystyle{ecta}
\bibliography{DSM}

\newpage

\appendix

\section{Calculation example} 
\label{sec:calc_example}
\FloatBarrier

The following illustrates how answers to the debt aversion survey module are translated into $\hat{\gamma}$ the predicted structural estimate of debt aversion.

\begin{figure}[h]
    \begin{center}
        \fboxsep=12pt
        \fbox{
        \begin{minipage}{0.8\textwidth}
        Please rate the following statement.\bigbreak
        
        \hspace{12pt}\textit{\textbf{Debt is an integral part of today's life.}}\vspace{-12pt}
        
            \begin{center}
            \begin{tabular}{c c c c c c}
                 {\small strongly agree \hspace{6pt} $\square$}
                 & {\small $\square$}
                 & {\small $\square$}
                 & {\small $\square$}
                 & {\small $\boxtimes$}
                 & {\small $\square$ \hspace{6pt} strongly disagree }
            \end{tabular}
            \end{center}
        
            
            
        \begin{center} --- \end{center}
        
        What do you think how does the average participant in this survey/experiment rate the following statement?\bigbreak
        
        \hspace{12pt}\textit{\textbf{There is no excuse for borrowing money.}}\vspace{-12pt}
        
            \begin{center}
            \begin{tabular}{c c c c c c}
                 {\small strongly agree \hspace{6pt} $\square$}
                 & {\small $\boxtimes$}
                 & {\small $\square$}
                 & {\small $\square$}
                 & {\small $\square$}
                 & {\small $\square$ \hspace{6pt} strongly disagree }\\
            \end{tabular}
            \end{center}
        
        \end{minipage}
        }
    \end{center}

\caption{Debt aversion survey module answered by a hypothetical respondent}
\label{fig.example_questionnaire}
\end{figure}

Let's consider the responses to the debt aversion suvery module by a hypothetical person as depicted in Figure~\ref{fig.example_questionnaire}. In this particular case, the predicted debt aversion parameter, $\hat{\gamma}$, can be calculated as

\begin{flalign*}
    \hat{\gamma} = &1.0694 + \\
    &0.0045\times (5) - 0.0067\times (2)\\
    = &1.0785
\end{flalign*}

where the numbers in brackets are the answers given to the survey items coded into numbers 1-6 from the left-most to the right-most answer increment, respectively.

We recommend implementing the survey module with the presented 6-point Likert scales, as this format is underlying all the validation procedures. However, should there be reasons to use Likert scales with more or less increments, the predicted debt aversion parameter $\hat{\gamma}$, can be calculated from Likert scales with minimum $l$ and maximum $h$ by transforming the answers $x$ to the 6-point scale. To this end, one needs to replace the numbers in brackets by:
 
 \begin{align}
    \left(\frac{x-l}{h-l} \times 5+1\right)
\end{align}

\FloatBarrier
\newpage

\section{Theoretical Framework}
\label{sec:model}

In \cite{meissner2022}, agents choose between intertemporal prospects that are defined over streams of monetary gains or losses in up to two periods. $\boldsymbol{x}=(x_t,x_T)$ denotes a stream of payments that offers $x_t$ at time $t$, and $x_T$ at time $T$, where $0\leq t < T$. $X=(\boldsymbol{x}_1,p_1;\boldsymbol{x}_2,p_2;...;\boldsymbol{x}_N,p_N)$ denotes an intertemporal prospect, that gives the payment stream $\boldsymbol{x}_n$ with probability $p_n$. Intertemporal utility is written as:

\begin{equation*}
\label{eq:U}
    U(X)=\mathbb{E}\left[\phi(t)v(x_t)+\phi \left(T\right)v(x_T) - \mathbbm{1}_{debt} c\left(\boldsymbol{x}\right)\right]
\end{equation*}

where $v(x_t)$ denotes atemporal utility of monetary gains and losses at time $t$. Agents discount future gains and losses with the discount function $\phi$. 

\emph{Saving contracts} are payment streams characterized by $x_t<0$ and $x_T>0$.  Inversely, \emph{debt contracts} are payment streams characterized by $x_t>0$ and $x_T<0$. Cost of being in debt $c(\boldsymbol x)$ is only incurred for debt contracts:

\begin{equation*}
    \mathbbm{1}_{debt}=\begin{cases}
    1 & \mbox{if } x_t > 0 \mbox{ and } x_T <0\\
    0 & \mbox{ otherwise.}
    \end{cases}
\end{equation*}

The specific functional forms used to estimate preference parameters are identical to \cite{meissner2022}. Gains and losses of money are evaluated relative to a reference point $(x=0)$:

\begin{align}
v(x)=
\begin{cases}
u(x) & \mbox{if } x\geq 0\\
-\lambda u(-x) & \mbox{if } x<0,\\
\end{cases}
\end{align}

Utility in gains and losses is given by: 

\begin{align}
    u(x)=\frac{(x+\varepsilon)^{1-\alpha}-\varepsilon^{1-\alpha}}{1-\alpha}
\end{align}

The discount funcion is exponential:

\begin{align}
    \phi(\tau) = \frac{1}{(1+\delta)^{\tau}}   
\end{align}

Finally, the cost of being in debt is modelled as:

\begin{align}
   c(\boldsymbol{x})=(1-\gamma)\phi(T)v(x_T)
\end{align}

Here, $\gamma$ is the parameter of debt aversion. A parameter of $\gamma=1$ implies debt neutrality, $\gamma>1$ implies debt aversion and $\gamma<1$ implies debt affinity.

\section{Multiple Price Lists} \label{sec:mpls}
\FloatBarrier

{\def\arraystretch{0.75}
\begin{table}[htp]
\begin{tabularx}{\textwidth}{ 
    c
    @{\hspace{36pt}}
  >{\centering\arraybackslash}X 
  >{\centering\arraybackslash}X 
  }
\toprule

\textbf{Choice} 
& \textbf{Option A}
& \textbf{Option B}\\
\hline

\footnotesize{1}
& \footnotesize{Receive an amount of \euro{18.2} today}
& \footnotesize{Receive an amount of  \euro{18.0} in 4 weeks}\\

\footnotesize{2}
& \footnotesize{Receive an amount of \euro{18.0} today}
& \footnotesize{Receive an amount of  \euro{18.0} in 4 weeks}\\

\footnotesize{3} 
& \footnotesize{Receive an amount of  \euro{17.8} today} 
& \footnotesize{Receive an amount of  \euro{18.0} in 4 weeks}\\

\footnotesize{4}
& \footnotesize{Receive an amount of  \euro{17.3} today}
& \footnotesize{Receive an amount of  \euro{18.0} in 4 weeks}\\

\footnotesize{5}
& \footnotesize{Receive an amount of  \euro{16.8} today}
& \footnotesize{Receive an amount of  \euro{18.0} in 4 weeks}\\

\footnotesize{6}
& \footnotesize{Receive an amount of  \euro{16.0} today}
& \footnotesize{Receive an amount of  \euro{18.0} in 4 weeks}\\

\footnotesize{7}
& \footnotesize{Receive an amount of  \euro{14.0} today}
& \footnotesize{Receive an amount of  \euro{18.0} in 4 weeks}\\

\footnotesize{8}
& \footnotesize{Receive an amount of  \euro{12.0} today}
& \footnotesize{Receive an amount of  \euro{18.0} in 4 weeks}\\

\footnotesize{9}
& \footnotesize{Receive an amount of  \euro{10.0} today}
& \footnotesize{Receive an amount of  \euro{18.0} in 4 weeks}\\

\footnotesize{10}
& \footnotesize{Receive an amount of  \euro{8.0} today}
& \footnotesize{Receive an amount of  \euro{18.0} in 4 weeks}\\

\bottomrule
\end{tabularx}
\caption{Multiple price list of intertemporal choices (MPL1)}
\label{tab:MPLI}
\end{table}

\begin{table}[htp]
\begin{tabularx}{\textwidth}{ 
c
@{\hspace{36pt}}
C{0.1cm} 
>{\centering\arraybackslash}X 
>{\centering\arraybackslash}X 
C{0.1cm} 
>{\centering\arraybackslash}X 
>{\centering\arraybackslash}X  
C{0.1cm}
}
\toprule
& & \multicolumn{2}{c}{\textbf{Option A}} & & \multicolumn{2}{c}{\textbf{Option B}} & \\
\cline{3-4} \cline{6-7}
\textbf{Choice}  & &  \makecell{\small{ Coin shows  } \\[-0.2cm] \small{ Heads }} & \makecell*{\small{ Coin shows  } \\[-0.2cm] \small{ Tails }} & & \makecell*{\small{ Coin shows  } \\[-0.2cm] \small{ Heads }} & \makecell*{\small{ Coin shows  } \\[-0.2cm] \small{ Tails }} & \\
\hline
\footnotesize{1} & & \footnotesize{\euro{30}  today} & \footnotesize{\euro{30}  today} & & \footnotesize{\euro{30}  today} & \footnotesize{\euro{1}  today} & \\
\footnotesize{2} & & \footnotesize{\euro{25}  today} & \footnotesize{\euro{25}  today} & & \footnotesize{\euro{30}  today} & \footnotesize{\euro{1}  today} & \\
\footnotesize{3} & & \footnotesize{\euro{20}  today} & \footnotesize{\euro{20}  today} & & \footnotesize{\euro{30} today} & \footnotesize{\euro{1}  today} & \\
\footnotesize{4} & & \footnotesize{\euro{17}  today} & \footnotesize{\euro{17}  today} & & \footnotesize{\euro{30}  today} & \footnotesize{\euro{1}  today} & \\
\footnotesize{5} & & \footnotesize{\euro{16}  today} & \footnotesize{\euro{16}  today} & & \footnotesize{\euro{30}  today} & \footnotesize{\euro{1}  today} & \\
\footnotesize{6} & & \footnotesize{\euro{15}  today} & \footnotesize{\euro{15}  today} & & \footnotesize{\euro{30}  today} & \footnotesize{\euro{1} today} & \\
\footnotesize{7} & &\footnotesize{\euro{12}  today} & \footnotesize{\euro{12}  today} & & \footnotesize{\euro{30}  today} & \footnotesize{\euro{1}  today} & \\
\footnotesize{8} & & \footnotesize{\euro{10}\  today} & \footnotesize{\euro{10}  today} & & \footnotesize{\euro{30}  today} & \footnotesize{\euro{1}  today} & \\
\footnotesize{9} & &\footnotesize{\euro{5}  today} & \footnotesize{\euro{5}  today} & & \footnotesize{\euro{30}  today} & \footnotesize{\euro{1}  today} & \\
\footnotesize{10} & & \footnotesize{\euro{1}  today} & \footnotesize{\euro{1}  today} & & \footnotesize{\euro{30}  today} & \footnotesize{\euro{1}  today} &\\
\bottomrule
\end{tabularx}
\caption{Multiple price list of certain payments vs. risky gambles (MPL2)}
\label{tab:MPLII}
\end{table}

\begin{table}[htp]
\begin{tabularx}{\textwidth}{ 
c
@{\hspace{36pt}}
C{0.1cm} 
>{\centering\arraybackslash}X 
>{\centering\arraybackslash}X 
C{0.1cm} 
>{\centering\arraybackslash}X 
>{\centering\arraybackslash}X  
C{0.1cm}
}
\toprule
& & \multicolumn{2}{c}{\textbf{Option A}} & & \multicolumn{2}{c}{\textbf{Option B}} & \\
\cline{3-4} \cline{6-7}
\textbf{Choice}  & &  \makecell{\small{ Coin shows  } \\[-0.2cm] \small{ Heads }} & \makecell*{\small{ Coin shows  } \\[-0.2cm] \small{ Tails }} & & \makecell*{\small{ Coin shows  } \\[-0.2cm] \small{ Heads }} & \makecell*{\small{ Coin shows  } \\[-0.2cm] \small{ Tails }} & \\
\hline
\footnotesize{1} & & \footnotesize{\euro{14}  today} & \footnotesize{\euro{17}  today} & & \footnotesize{\euro{17}  today} & \footnotesize{\euro{1} today} & \\
\footnotesize{2} & & \footnotesize{\euro{14}  today} & \footnotesize{\euro{17}  today} & & \footnotesize{\euro{20}  today} & \footnotesize{\euro{1} today} & \\
\footnotesize{3} & & \footnotesize{\euro{14}  today} & \footnotesize{\euro{17}  today} & & \footnotesize{\euro{25}  today} & \footnotesize{\euro{1} today} & \\
\footnotesize{4} & & \footnotesize{\euro{14}  today} & \footnotesize{\euro{17} today} & & \footnotesize{\euro{28}  today} & \footnotesize{\euro{1} today} & \\
\footnotesize{5} & & \footnotesize{\euro{14}  today} & \footnotesize{\euro{17}  today} & & \footnotesize{\euro{29}  today} & \footnotesize{\euro{1} today} & \\
\footnotesize{6} & & \footnotesize{\euro{14}  today} & \footnotesize{\euro{17}  today} & & \footnotesize{\euro{30}  today} & \footnotesize{\euro{2}  today} & \\
\footnotesize{7} & &\footnotesize{\euro{14} today} & \footnotesize{\euro{17}  today} & & \footnotesize{\euro{30}  today} & \footnotesize{\euro{3}  today} & \\
\footnotesize{8} & & \footnotesize{\euro{14}  today} & \footnotesize{\euro{17}  today} & & \footnotesize{\euro{32}  today} & \footnotesize{\euro{8}  today} & \\
\footnotesize{9} & &\footnotesize{\euro{14}  today} & \footnotesize{\euro{17}  today} & & \footnotesize{\euro{32}  today} & \footnotesize{\euro{10}  today} & \\
\footnotesize{10} & & \footnotesize{\euro{14}  today} & \footnotesize{\euro{17}  today} & & \footnotesize{\euro{32}  today} & \footnotesize{\euro{14}  today} & \\
\bottomrule
\end{tabularx}
\caption{Multiple price list of less risky vs. more risky gambles (MPL3)}
\label{tab:MPLIII}
\end{table}

\begin{table}[htp]
\begin{tabularx}{\textwidth}{
    c
    @{\hspace{36pt}}
    C{0.1cm} 
  >{\centering\arraybackslash}X 
    C{0.1cm} 
  >{\centering\arraybackslash}X 
    C{0.1cm}
    }
\toprule
 & & \multicolumn{3}{c}{\textbf{Early saving contracts}} & \\
\cline{3-5}
\textbf{Choice} & & \makecell{\small{ Session 1 } \\[-0.2cm] \footnotesize{ (today) }} & & \makecell{\small{ Session 2 } \\[-0.2cm] \footnotesize{ (in 4 weeks) }} & \\

\hline
\footnotesize{1} & & \footnotesize{Pay an amount of \euro{15}} & & \footnotesize{Receive an amount of  \euro{45}}\\
\footnotesize{2} & & \footnotesize{Pay an amount of \euro{15}} & & \footnotesize{Receive an amount of  \euro{40}}\\
\footnotesize{3} & & \footnotesize{Pay an amount of \euro{15}} & & \footnotesize{Receive an amount of  \euro{36}}\\
\footnotesize{4} & & \footnotesize{Pay an amount of \euro{15}} & & \footnotesize{Receive an amount of  \euro{34}}\\
\footnotesize{5} & & \footnotesize{Pay an amount of \euro{15}} & & \footnotesize{Receive an amount of  \euro{32}}\\
\footnotesize{6} & & \footnotesize{Pay an amount of \euro{15}} & & \footnotesize{Receive an amount of  \euro{30}}\\
\footnotesize{7} & & \footnotesize{Pay an amount of \euro{15}} & & \footnotesize{Receive an amount of  \euro{28}}\\
\footnotesize{8} & & \footnotesize{Pay an amount of \euro{15}} & & \footnotesize{Receive an amount of  \euro{26}}\\
\footnotesize{9} & & \footnotesize{Pay an amount of \euro{15}} & & \footnotesize{Receive an amount of  \euro{24}}\\
\footnotesize{10} & & \footnotesize{Pay an amount of \euro{15}} & & \footnotesize{Receive an amount of  \euro{22}}\\
\footnotesize{11} & & \footnotesize{Pay an amount of \euro{15}} & & \footnotesize{Receive an amount of  \euro{20}}\\
\footnotesize{12} & & \footnotesize{Pay an amount of \euro{15}} & & \footnotesize{Receive an amount of  \euro{18}}\\
\footnotesize{13} & & \footnotesize{Pay an amount of \euro{15}} & & \footnotesize{Receive an amount of  \euro{16}}\\
\footnotesize{14} & & \footnotesize{Pay an amount of \euro{15}} & & \footnotesize{Receive an amount of  \euro{14}}\\
\footnotesize{15} & & \footnotesize{Pay an amount of \euro{15}} & & \footnotesize{Receive an amount of  \euro{12}}\\

\bottomrule
\end{tabularx}
\caption{Multiple price list of 4-week saving contracts starting at Session 1 (MPL4)}
\label{tab:MPLIV}
\end{table}

\begin{table}[htp]
\begin{tabularx}{\textwidth}{
    c
    @{\hspace{36pt}}
    C{0.1cm} 
  >{\centering\arraybackslash}X 
    C{0.1cm} 
  >{\centering\arraybackslash}X 
    C{0.1cm}
    }
\toprule
 & & \multicolumn{3}{c}{\textbf{Late saving contracts}} & \\
\cline{3-5}
\textbf{Choice} & & \makecell{\small{ Session 2 } \\[-0.2cm] \footnotesize{ (in 4 weeks) }} & & \makecell{\small{ Session 3 } \\[-0.2cm] \footnotesize{ (in 8 weeks) }} & \\
\hline
\footnotesize{1} & & \footnotesize{Pay an amount of \euro{15}} & & \footnotesize{Receive an amount of  \euro{40}}\\
\footnotesize{2} & & \footnotesize{Pay an amount of \euro{15}} & & \footnotesize{Receive an amount of  \euro{35}}\\
\footnotesize{3} & & \footnotesize{Pay an amount of \euro{15}} & & \footnotesize{Receive an amount of  \euro{31}}\\
\footnotesize{4} & & \footnotesize{Pay an amount of \euro{15}} & & \footnotesize{Receive an amount of  \euro{29}}\\
\footnotesize{5} & & \footnotesize{Pay an amount of \euro{15}} & & \footnotesize{Receive an amount of  \euro{27}}\\
\footnotesize{6} & & \footnotesize{Pay an amount of \euro{15}} & & \footnotesize{Receive an amount of  \euro{25}}\\
\footnotesize{7} & & \footnotesize{Pay an amount of \euro{15}} & & \footnotesize{Receive an amount of  \euro{23}}\\
\footnotesize{8} & & \footnotesize{Pay an amount of \euro{15}} & & \footnotesize{Receive an amount of  \euro{21}}\\
\footnotesize{9} & & \footnotesize{Pay an amount of \euro{15}} & & \footnotesize{Receive an amount of  \euro{19}}\\
\footnotesize{10} & & \footnotesize{Pay an amount of \euro{15}} & & \footnotesize{Receive an amount of  \euro{17}}\\
\footnotesize{11} & & \footnotesize{Pay an amount of \euro{15}} & & \footnotesize{Receive an amount of  \euro{15}}\\
\footnotesize{12} & & \footnotesize{Pay an amount of \euro{15}} & & \footnotesize{Receive an amount of  \euro{13}}\\
\footnotesize{13} & & \footnotesize{Pay an amount of \euro{15}} & & \footnotesize{Receive an amount of  \euro{11}}\\
\footnotesize{14} & & \footnotesize{Pay an amount of \euro{15}} & & \footnotesize{Receive an amount of  \euro{9}}\\
\footnotesize{15} & & \footnotesize{Pay an amount of \euro{15}} & & \footnotesize{Receive an amount of  \euro{7}}\\

\bottomrule
\end{tabularx}
\caption{Multiple price list of 4-week saving contracts starting at Session 2 (MPL5)}
\label{tab:MPLV}
\end{table}

\begin{table}[htp]
\begin{tabularx}{\textwidth}{
    c
    @{\hspace{36pt}}
    C{0.1cm} 
  >{\centering\arraybackslash}X 
    C{0.1cm} 
  >{\centering\arraybackslash}X 
    C{0.1cm}
    }
\toprule
 & & \multicolumn{3}{c}{\textbf{Early debt contracts}} & \\
\cline{3-5}
\textbf{Choice} & & \makecell{\small{ Session 1 } \\[-0.2cm] \footnotesize{ (today) }} & & \makecell{\small{ Session 2 } \\[-0.2cm] \footnotesize{ (in 4 weeks) }} & \\
\hline
\footnotesize{1} & & \footnotesize{Receive an amount of \euro{31}} & & \footnotesize{Pay an amount of  \euro{15}}\\
\footnotesize{2} & & \footnotesize{Receive an amount of \euro{27}} & & \footnotesize{Pay an amount of  \euro{15}}\\
\footnotesize{3} & & \footnotesize{Receive an amount of \euro{24}} & & \footnotesize{Pay an amount of  \euro{15}}\\
\footnotesize{4} & & \footnotesize{Receive an amount of \euro{21}} & & \footnotesize{Pay an amount of  \euro{15}}\\
\footnotesize{5} & & \footnotesize{Receive an amount of \euro{19}} & & \footnotesize{Pay an amount of  \euro{15}}\\
\footnotesize{6} & & \footnotesize{Receive an amount of \euro{17}} & & \footnotesize{Pay an amount of  \euro{15}}\\
\footnotesize{7} & & \footnotesize{Receive an amount of \euro{16}} & & \footnotesize{Pay an amount of  \euro{15}}\\
\footnotesize{8} & & \footnotesize{Receive an amount of \euro{15}} & & \footnotesize{Pay an amount of  \euro{15}}\\
\footnotesize{9} & & \footnotesize{Receive an amount of \euro{14}} & & \footnotesize{Pay an amount of  \euro{15}}\\
\footnotesize{10} & & \footnotesize{Receive an amount of \euro{13}} & & \footnotesize{Pay an amount of  \euro{15}}\\
\footnotesize{11} & & \footnotesize{Receive an amount of \euro{11}} & & \footnotesize{Pay an amount of  \euro{15}}\\
\footnotesize{12} & & \footnotesize{Receive an amount of \euro{9}} & & \footnotesize{Pay an amount of  \euro{15}}\\
\footnotesize{13} & & \footnotesize{Receive an amount of \euro{7}} & & \footnotesize{Pay an amount of  \euro{15}}\\
\footnotesize{14} & & \footnotesize{Receive an amount of \euro{5}} & & \footnotesize{Pay an amount of  \euro{15}}\\
\footnotesize{15} & & \footnotesize{Receive an amount of \euro{3}} & & \footnotesize{Pay an amount of  \euro{15}}\\

\bottomrule
\end{tabularx}
\caption{Multiple price list of 4-week debt contracts starting at Session 1 (MPL6)}
\label{tab:MPLVI}
\end{table}

\begin{table}[htp]
\begin{tabularx}{\textwidth}{
    c
    @{\hspace{36pt}}
    C{0.1cm} 
  >{\centering\arraybackslash}X 
    C{0.1cm} 
  >{\centering\arraybackslash}X 
    C{0.1cm}
    }
\toprule
 & & \multicolumn{3}{c}{\textbf{Late debt contracts}} & \\
\cline{3-5}
\textbf{Choice} & & \makecell{\small{ Session 2 } \\[-0.2cm] \footnotesize{ (in 4 weeks) }} & & \makecell{\small{ Session 3 } \\[-0.2cm] \footnotesize{ (in 8 weeks) }} & \\
\hline
\footnotesize{1} & & \footnotesize{Receive an amount of \euro{33}} & & \footnotesize{Pay an amount of  \euro{15}}\\
\footnotesize{2} & & \footnotesize{Receive an amount of \euro{30}} & & \footnotesize{Pay an amount of  \euro{15}}\\
\footnotesize{3} & & \footnotesize{Receive an amount of \euro{27}} & & \footnotesize{Pay an amount of  \euro{15}}\\
\footnotesize{4} & & \footnotesize{Receive an amount of \euro{24}} & & \footnotesize{Pay an amount of  \euro{15}}\\
\footnotesize{5} & & \footnotesize{Receive an amount of \euro{22}} & & \footnotesize{Pay an amount of  \euro{15}}\\
\footnotesize{6} & & \footnotesize{Receive an amount of \euro{20}} & & \footnotesize{Pay an amount of  \euro{15}}\\
\footnotesize{7} & & \footnotesize{Receive an amount of \euro{18}} & & \footnotesize{Pay an amount of  \euro{15}}\\
\footnotesize{8} & & \footnotesize{Receive an amount of \euro{16}} & & \footnotesize{Pay an amount of  \euro{15}}\\
\footnotesize{9} & & \footnotesize{Receive an amount of \euro{15}} & & \footnotesize{Pay an amount of  \euro{15}}\\
\footnotesize{10} & & \footnotesize{Receive an amount of \euro{14}} & & \footnotesize{Pay an amount of  \euro{15}}\\
\footnotesize{11} & & \footnotesize{Receive an amount of \euro{12}} & & \footnotesize{Pay an amount of  \euro{15}}\\
\footnotesize{12} & & \footnotesize{Receive an amount of \euro{10}} & & \footnotesize{Pay an amount of  \euro{15}}\\
\footnotesize{13} & & \footnotesize{Receive an amount of \euro{8}} & & \footnotesize{Pay an amount of  \euro{15}}\\
\footnotesize{14} & & \footnotesize{Receive an amount of \euro{6}} & & \footnotesize{Pay an amount of  \euro{15}}\\
\footnotesize{15} & & \footnotesize{Receive an amount of \euro{3}} & & \footnotesize{Pay an amount of  \euro{15}}\\

\bottomrule
\end{tabularx}
\caption{Multiple price list of 4-week debt contracts starting at Session 2 (MPL7)}
\label{tab:MPLVII}
\end{table}

}
\FloatBarrier
\newpage

 \newpage
 \section{Debt Aversion Survey Module} \label{sec:DSM_a}
\FloatBarrier 

 \subsection{Pool of Debt Survey Items} \label{sec:DSM_Pool_a}
 
\begin{xltabular}{\textwidth}{c L{8cm} C{2cm} C{3cm}}
\label{table:debtsurvey_all}\\

\toprule
\textbf{No.} & \textbf{Survey Item} & \textbf{Scale} & \textbf{Reference} \\
\midrule
\endfirsthead

\textbf{No.} & \textbf{Survey Item} & \textbf{Scale} & \textbf{Reference} \\
\midrule
\endhead
\bottomrule
\multicolumn{2}{r}{\footnotesize(Continued on next page)}
\endfoot
\bottomrule
\caption{Pool of Debt Survey Items. All Likert scales follow a 6-point format, items without reference were created by the authors.}\\
\endlastfoot
\multicolumn{2}{l}{\footnotesize{\textit{Usage}}} & & \\
\cline{1-2}
\footnotesize{1} &\footnotesize{Did you ever use overdraft on your bank account?} & \footnotesize{yes/no} & \scriptsize{-}\\
\footnotesize{2} &\footnotesize{Do you use credit cards?} & \footnotesize{yes/no} & \scriptsize{\citep{RN147}} \\
\footnotesize{3} &\footnotesize{In total, how many credit cards with different accounts do you use?} & \footnotesize{categorical 0~to~$>5$} & \scriptsize{-}\\
\footnotesize{4} &\footnotesize{If you have a credit card balance. Do you usually pay it off each month?} & \footnotesize{yes/no} & \scriptsize{\citep{RN147}} \\
\footnotesize{5} &\footnotesize{How would you categorize your access to loans/credits/capital?} & \footnotesize{Likert} & \scriptsize{-}\\
\footnotesize{6} &\footnotesize{Did you ever take out a loan at a bank?} & \footnotesize{yes/no} & \scriptsize{-}\\
\footnotesize{7} &\footnotesize{Do you owe money in student loans?} & \footnotesize{yes/no} & \scriptsize{\citep{RN147}}\\
\footnotesize{8} &\footnotesize{In total, what is your best guess of your outstanding debt as of today in \euro ? (including informal loans, family, friends, etc.)} & \footnotesize{integer} & \scriptsize{-}\\
\footnotesize{9} &\footnotesize{How certain are you about your guess on your overall outstanding debt?} &  \footnotesize{Likert} & \scriptsize{-}\\
\footnotesize{10} &\footnotesize{Does your current level of debt burden you?} & \footnotesize{Likert} & \scriptsize{\citep{RN147}}\\
\footnotesize{11} &\footnotesize{In total, what is your best guess of your savings as of today in  \euro ?} & \footnotesize{integer} & \scriptsize{-}\\
\footnotesize{12} &\footnotesize{How certain are you about your guess on your overall savings?} &  \footnotesize{Likert} & \scriptsize{-}\\

\multicolumn{4}{l}{\footnotesize{\textit{Appropriateness: ``Please rate the following statements"}}} \\
\cline{1-2}
\footnotesize{13} &\footnotesize{It is okay to accrue debt for living the style you desire.} & \footnotesize{Likert} & \scriptsize{\citep{RN164}}\\
\footnotesize{14} &\footnotesize{It is okay to be in debt if you know you can pay it off.} & \footnotesize{Likert} & \scriptsize{\citep{RN165}}\\
\footnotesize{15} &\footnotesize{It is ok to borrow money to pay for necessities (e.g. food, rent, utilities).} & \footnotesize{Likert} & \scriptsize{\multirow{8}{3cm}{adapted based on: \citep{RN163, RN165, RN149, RN171}}}\\
\footnotesize{16} &\footnotesize{It is ok  to borrow money to pay for essential purchases (e.g. car, housing, appliances).} & \footnotesize{Likert} & \\
\footnotesize{17} &\footnotesize{It is ok to borrow money to finance investments (e.g. tertiary education, starting a business, solar panels).} & \footnotesize{Likert} & \\
\footnotesize{18} &\footnotesize{It is ok to borrow money to pay for luxuries (e.g. expensive holiday, status symbols).} & \footnotesize{Likert} & \\
\footnotesize{19} &\footnotesize{Students should take the maximum permissible student debt (loans/overdraft, etc.).} & \footnotesize{Likert} & \scriptsize{\citep{RN164}}\\
\footnotesize{20} &\footnotesize{Debt is an integral part of today's life.} & \footnotesize{Likert} & \scriptsize{\citep{RN163,RN165,RN164}}\\
\footnotesize{21} &\footnotesize{Reducing/controlling debt leads to a better quality of life.} & \footnotesize{Likert} & \scriptsize{\citep{RN164}}\\
\footnotesize{22} &\footnotesize{Reducing/controlling debt leads to greater success.} & \footnotesize{Likert} & \scriptsize{\citep{RN164}}\\
\footnotesize{23} &\footnotesize{Reducing/controlling debt leads to feeling a sense of achievement.} & \footnotesize{Likert} & \scriptsize{\citep{RN164}}\\
\footnotesize{24} &\footnotesize{Reducing/controlling debt leads to a feeling that you are fitting in with friends.} & \footnotesize{Likert} & \scriptsize{\citep{RN164}}\\
\footnotesize{25} &\footnotesize{Reducing/controlling debt leads to being perceived as boring.} & \footnotesize{Likert} & \scriptsize{\citep{RN164}}\\
\footnotesize{26} &\footnotesize{Reducing/controlling debt leads to being perceived as tight.} & \footnotesize{Likert} & \scriptsize{\citep{RN164}}\\
\footnotesize{27} &\footnotesize{Reducing/controlling debt leads to enjoying yourself less.} & \footnotesize{Likert} & \scriptsize{\citep{RN164}}\\

\footnotesize{28} &\footnotesize{Once you are in debt it is very difficult to get out of it.} & \footnotesize{Likert} & \scriptsize{\citep{RN163, RN165}}\\
\footnotesize{29} &\footnotesize{Owing money is basically wrong.} & \footnotesize{Likert} & \scriptsize{\citep{RN165,RN170}}\\
\footnotesize{30} &\footnotesize{You should always save up first before buying something.} & \footnotesize{Likert} & \scriptsize{\citep{RN163,RN165,RN170}}\\
\footnotesize{31} &\footnotesize{There is no excuse for borrowing money.} & \footnotesize{Likert} & \scriptsize{\citep{RN163,RN165,RN170}}\\

\footnotesize{32} &\footnotesize{Borrowing money for tertiary education is a good investment.} & \footnotesize{Likert} & \scriptsize{\citep{RN165}}\\

\footnotesize{33} &\footnotesize{You should rather restrict your lifestyle than go into debt.} & \footnotesize{Likert} & \scriptsize{-}\\

\multicolumn{4}{l}{\footnotesize{\textit{Personality: ``Please rate the following statements"}}} \\
\cline{1-2}

\footnotesize{34} &\footnotesize{I like to pay my debts as soon as possible.} & \footnotesize{Likert} & \scriptsize{\citep{RN168}}\\
\footnotesize{35} &\footnotesize{I prefer to delay paying my debts if possible, even if it means paying more in total.} & \footnotesize{Likert} & \scriptsize{\citep{RN168}}\\
\footnotesize{36} &\footnotesize{Having debts makes me feel uncomfortable.} & \footnotesize{Likert} & \scriptsize{\citep{RN168}}\\
\footnotesize{37} &\footnotesize{Having debt doesn't bother me.} & \footnotesize{Likert} & \scriptsize{\citep{RN168}}\\

\footnotesize{38} &\footnotesize{I dislike borrowing money.} & \footnotesize{Likert} & \scriptsize{\citep{schleich2021}}\\
\footnotesize{39} &\footnotesize{I feel OK borrowing money for `essential' purchases e.g. cars, appliances, mortgage.} & \footnotesize{Likert} & \scriptsize{\citep{schleich2021}}\\
\footnotesize{40} &\footnotesize{I enjoy being able to borrow money to buy things I like, and to pay for things I cannot afford.} & \footnotesize{Likert} & \scriptsize{\citep{schleich2021}}\\

\footnotesize{41} &\footnotesize{I would rather be in debt than change my lifestyle.} & \footnotesize{Likert} & \scriptsize{\citep{RN165}}\\

\footnotesize{42} &\footnotesize{If I had to make an unexpected expenditure today of 500~\euro\ or more, I would use a credit card/borrow from a financial institution, family or friends.} & \footnotesize{Likert} & \scriptsize{\citep{RN147}}\\
\footnotesize{43} &\footnotesize{If I had to make an unexpected expenditure today of 5000~\euro\ or more, I would use a credit card/borrow from a financial institution, family or friends.} & \footnotesize{Likert} & \scriptsize{\citep{RN147}}\\

\footnotesize{44} &\footnotesize{I like saving money.} & \footnotesize{Likert} & \scriptsize{-}\\

\multicolumn{2}{l}{\footnotesize{\textit{Norms, i.e. second order beliefs \citep{RN487}\footnote{``What do you think, how does the average participant in this experiment rate the following statements on borrowing money?"}}}} & &\\
\cline{1-2}

\footnotesize{45} &\footnotesize{It is okay to accrue debt for living the style you desire.} & \footnotesize{Likert} & \scriptsize{\citep{RN164}}\\
\footnotesize{46} &\footnotesize{Students should take the maximum permissible student debt (loans/overdraft, etc.).} & \footnotesize{Likert} & \scriptsize{\citep{RN164}}\\
\footnotesize{47} &\footnotesize{Debt is an integral part of today's life.} & \footnotesize{Likert} & \scriptsize{\citep{RN163, RN165, RN164}}\\

\footnotesize{48} &\footnotesize{Once you are in debt it is very difficult to get out of it.} & \footnotesize{Likert} & \scriptsize{\citep{RN163, RN165}}\\

\footnotesize{49} &\footnotesize{Owing money is basically wrong.} & \footnotesize{Likert} & \scriptsize{\citep{RN165, RN170}}\\

\footnotesize{50} &\footnotesize{You should always save up first before buying something.} & \footnotesize{Likert} & \scriptsize{\citep{RN163, RN165, RN170}}\\
\footnotesize{51} &\footnotesize{There is no excuse for borrowing money.} & \footnotesize{Likert} & \scriptsize{\citep{RN163, RN165, RN170}}\\

\footnotesize{52} &\footnotesize{It is okay to be in debt if you know you can pay it off.} & \footnotesize{Likert} & \scriptsize{\citep{RN165}}\\
\footnotesize{53} &\footnotesize{Borrowing money for tertiary education is a good investment.} & \footnotesize{Likert} & \scriptsize{\citep{RN165}}\\

\footnotesize{54} &\footnotesize{You should rather restrict your lifestyle than go into debt.} & \footnotesize{Likert} & \scriptsize{-}\\

\end{xltabular}

 \subsection{Hypothetical Debt Contracts} \label{sec:DSM_HDC_a}
The multiple price list on hypothetical debt contracts contains a total of 15 decisions. Aiming at brevity of the debt aversion survey module, its implementation and validity has been tested using the staircase method. Thus respondents effectively only see and make four yes/no choices. Participants are asked whether they would, hypothetically, accept four financial debt contracts, under which they receive 100\euro\ today, with the obligation to repay between 60 to 140\euro\ in six months:
 
 \begin{center}
 \textit{Imagine your bank offered you a debt contract. Under this contract you receive 100\euro\ from your bank today and have to pay back XX\euro\ in 6 months. Please assume that you must pay the full amount you owe to the bank on time.\\[0.3cm]
 Would you accept such a contract?}
 \end{center}
 
\begin{figure}
\tikzstyle{level 1}=[level distance=2cm, sibling distance=8cm]
\tikzstyle{level 2}=[level distance=3cm, sibling distance=4cm]
\tikzstyle{level 3}=[level distance=4cm, sibling distance=1.5cm]
\tikzstyle{level 4}=[level distance=4cm, sibling distance=0.5cm]

\tikzstyle{bag} = [text width=6em, text centered]
\tikzstyle{end} = [circle, minimum width=6pt,fill, inner sep=0pt]

\begin{tikzpicture}[grow=right, sloped]
\node[bag] {100; -100}
    child {
        node[bag] {100; -90}        
            child {
            node[bag] {100; -75}       
            		child {
                	node[bag] {100; -60}
                 		child{
                 		node{SP=1}
                 		edge from parent
                		node[above] {}
                		node[below]  {not accept}
                		}
                		child{
                 		node{SP=2}
                 		edge from parent
                		node[above] {accept}
                		node[below]  {}
                		}
                	edge from parent
                	node[above] {}
                	node[below]  {not accept}
           		}
           		child{
           	 	node[bag] {100; -85}
                 		child{
                 		node{SP=3}
                 		edge from parent
                		node[above] {}
                		node[below]  {not accept}
                		}
                		child{
                 		node{SP=4}
                 		edge from parent
                		node[above] {accept}
                		node[below]  {}
                		}
                	edge from parent
                	node[above] {accept}
                	node[below]  {}
           	 	}
            edge from parent      
        	   node[above] {}
        	   node[below]  {not accept}
            }
            child {
            node[bag] {100; -97}        
            		child {
                	node[bag] {100; -95}
                 		child{
                 		node{SP=5}
                 		edge from parent
                		node[above] {}
                		node[below]  {not accept}
                		}
                		child{
                 		node{SP=6}
                 		edge from parent
                		node[above] {accept}
                		node[below]  {}
                		}
                	edge from parent
                	node[above] {}
                	node[below]  {not accept}
           		}
           		child{
           	 	node[bag] {100; -99}
                 		child{
                 		node{SP=7}
                 		edge from parent
                		node[above] {}
                		node[below]  {not accept}
                		}
                		child{
                 		node{SP=8}
                 		edge from parent
                		node[above] {accept}
                		node[below]  {}
                		}
            		edge from parent      
    				node[above] {accept}
    				node[below]  {}
    				}
    		   edge from parent      
        	   node[above] {accept}
        	   node[below]  {}
            	}
    edge from parent      
    node[above] {}
    node[below]  {not accept}
    }
    child {
        node[bag] {100; -110}        
            child {
            node[bag] {100; -103}
           child {
                	node[bag] {100; -101}
                 		child{
                 		node{SP=9}
                 		edge from parent
                		node[above] {}
                		node[below]  {not accept}
                		}
                		child{
                 		node{SP=10}
                 		edge from parent
                		node[above] {accept}
                		node[below]  {}
                		}
                	edge from parent
                	node[above] {}
                	node[below]  {not accept}
           		}
           		child{
           	 	node[bag] {100; -105}
                 		child{
                 		node{SP=11}
                 		edge from parent
                		node[above] {}
                		node[below]  {not accept}
                		}
                		child{
                 		node{SP=12}
                 		edge from parent
                		node[above] {accept}
                		node[below]  {}
                		}
            		edge from parent      
    				node[above] {accept}
    				node[below]  {}
    				}
           	edge from parent
             	node[above] {}
            	node[below]  {not accept}
            	}
            child {
            node[bag] {100; -125}     
            		child {
                	node[bag] {100; -115}
                 		child{
                 		node{SP=13}
                 		edge from parent
                		node[above] {}
                		node[below]  {not accept}
                		}
                		child{
                 		node{SP=14}
                 		edge from parent
                		node[above] {accept}
                		node[below]  {}
                		}
                	edge from parent
                	node[above] {}
                	node[below]  {not accept}
           		}
           		child{
           	 	node[bag] {100; -140}
                 		child{
                 		node{SP=15}
                 		edge from parent
                		node[above] {}
                		node[below]  {not accept}
                		}
                		child{
                 		node{SP=16}
                 		edge from parent
                		node[above] {accept}
                		node[below]  {}
                		}
            		edge from parent      
    				node[above] {accept}
    				node[below]  {}
    				}
          	edge from parent
         	node[above] {accept}
            	node[below]  {}   
            	}
    edge from parent
   	node[above] {accept}
   	node[below]  {}   
    };
\end{tikzpicture}
\caption{Staircase elicitation of choices on hypothetical debt contracts}
\label{fig.DSM_STC_a}
\end{figure}
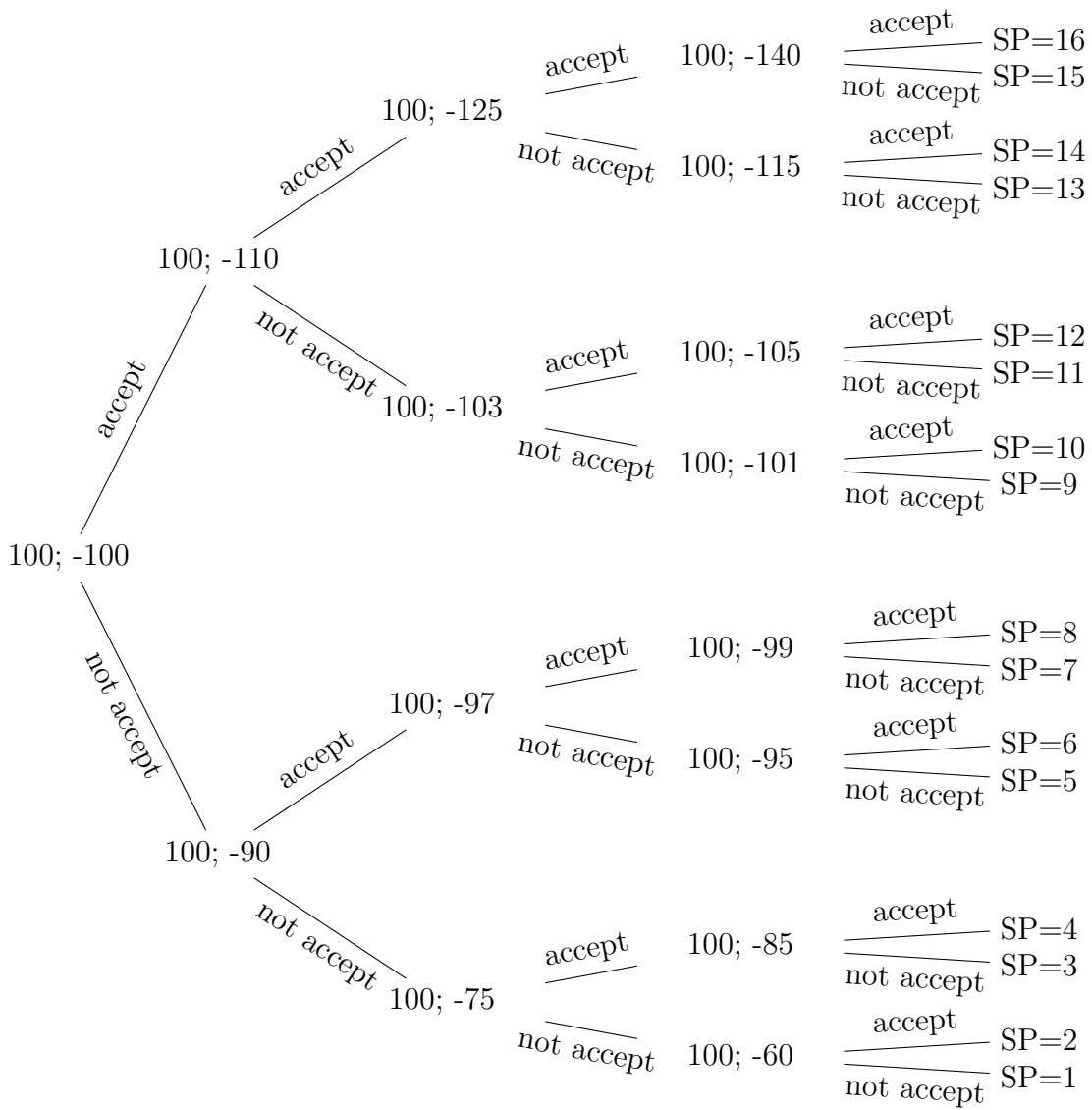

\noindent Figure~\ref{fig.DSM_STC_a} illustrates all 15 choices of the MPL. Nodes depict a hypothetical debt contract. Branches depict the available choices. Based on the choice in a specific node the path through the staircase is determined. The fixed, positive amounts of 100\euro\ indicates the hypothetical amount to be received today and the node/contract-specific negative amounts indicate the respective hypothetical amount of repayment in six months (to be inserted in for XX in the above mentioned question text). Respondents start at the left-most node and work their way through four questions until reaching a end-point at the right side. The label in the end-point states the switchpoint (SP) associated with the given choice path.

\FloatBarrier

\end{document}